\documentclass[journal=nalefd,manuscript=letter]{achemso}
\usepackage{amsmath}
\usepackage{physics}
\usepackage[super]{natbib}
\makeatletter
\renewcommand\@biblabel[1]{#1}
\makeatother
\NewDocumentCommand{\parencite}{m}{\begingroup\bibpunct{}{}{,}{n}{}{,}\cite{#1}\endgroup}
\usepackage{amsfonts}
\usepackage{mathtools}
\usepackage{bm}
\usepackage{graphicx}
\usepackage{hyperref}
\usepackage{pifont}
\usepackage{xcolor}
\usepackage{tabularx}
\usepackage{comment}
\allowdisplaybreaks
\author{MingRui Lai}
\affiliation{Department of Physics, National University of Singapore, 2 Science Drive 3, Singapore 117551}
\alsoaffiliation{Integrative Sciences and Engineering Programme, NUS Graduate School, National University of Singapore}
\alsoaffiliation{Centre for Advanced 2D Materials, National University of Singapore, 6 Science Drive 2, Singapore 117546}
\author{Fengyuan Xuan}
\alsoaffiliation{Department of Physics, National University of Singapore, 2 Science Drive 3, Singapore 117551}
\alsoaffiliation{Centre for Advanced 2D Materials, National University of Singapore, 6 Science Drive 2, Singapore 117546}

\author{Su Ying Quek}
\email{phyqsy@nus.edu.sg}
\alsoaffiliation{Department of Physics, National University of Singapore, 2 Science Drive 3, Singapore 117551}
\alsoaffiliation{Centre for Advanced 2D Materials, National University of Singapore, 6 Science Drive 2, Singapore 117546}
\alsoaffiliation{Integrative Sciences and Engineering Programme, NUS Graduate School, National University of Singapore}
\alsoaffiliation{Department of Materials Science and Engineering, National University of Singapore}
\title{Quantum Geometric Advantage of the Correlated Exciton State in Non-linear Optics }
\date{\today}
\begin{document}
\begin{abstract}
The concept of quantum geometry for single-particle states has revolutionized our interpretation of several emergent properties in condensed matter. However, a description of the quantum geometry for interacting particles and an understanding of its implications are lacking. Here, we show that inherent in the non-linear optical response is a quantum geometry of the correlated electron-hole state (exciton) that arises from the interplay between geometry and interactions - distinct from the quantum geometric properties of the individual electron or hole states. We demonstrate using first principles calculations that this quantum many-body geometry significantly enhances the non-linear optical response in systems with strong excitonic effects. In the case of shift currents, the quantum many-body geometric term arises in the many-body shift vector and can be interpreted as a many-body analogue of the Berry phase. This work lays the foundation to study the quantum geometry of correlated states in experimentally observable settings.
\end{abstract}
Quantum geometry, the description of quantum systems using concepts from geometry, is of fundamental importance in emergent quantum phenomena such as the macroscopic electronic polarization~\cite{king1993theory}, the integer quantum Hall effect~\cite{thouless1982quantized}, the anomalous Hall effect~\cite{karplus1954hall} and non-linear phenomena~\cite{sodemann2015quantum}. Examples of quantum geometric quantities include the Berry phase~\cite{berry1984quantal}, a geometric phase accumulated by the evolution of quantum states over a closed path in parameter space and the Berry connection, whose integral gives the Berry phase. The discussion of quantum geometry within these contexts have been confined to single-particle states. \\

The single-particle approximation can fail in systems with strong correlations, and it is important to know how quantum geometric concepts can be applied in the presence of interactions. Is the quantum geometry of an interacting system simply a superposition of quantum geometric properties of single particle states, or are there additional effects arising from an interplay between quantum geometry and interactions?  \\

\begin{figure}
\centering
\includegraphics[width=0.5\textwidth]{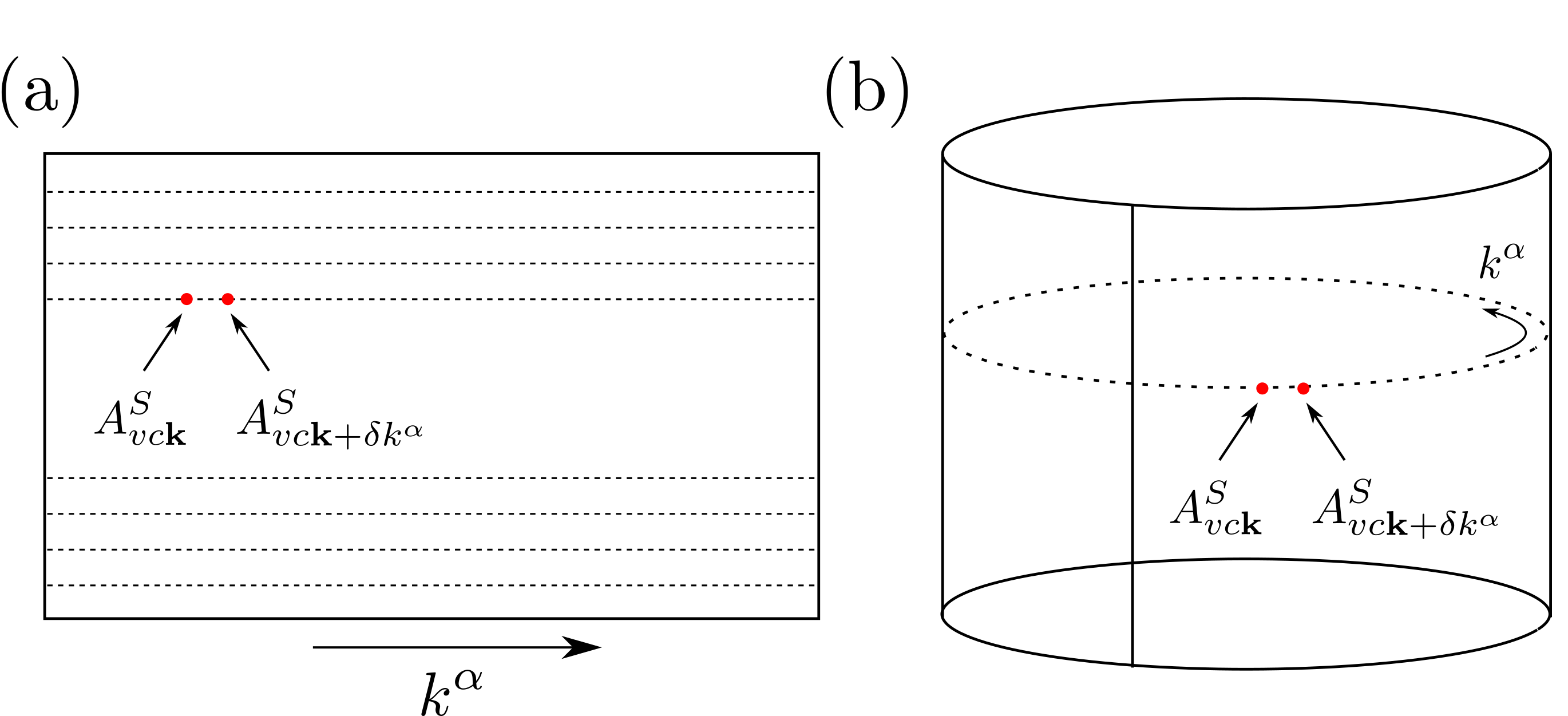}
    \caption{(a) Paths in the two dimension Brillouin zone along which line integrals of the geometric term are carried out. The geometric term sums the contribution of the product of the exciton envelope function at two neighbouring $k$-points along the straight dotted lines across the Brillouin zone in the direction of $k^\alpha$. (b) The same Brillouin zone in (a) but folded onto its edge into a cylinder, highlighting its periodicity. The integral is now an integral along a closed loop. This geometric term can thus be viewed as sums along these closed paths, accumulating an overall phase.}
    \label{geometric}
\end{figure}

Excitons are correlated electron-hole states that arise from neutral single excitations; excitons are fundamental to the description of optical properties~\cite{rohlfing2000electron} and the development of modern optics~\cite{ciarrocchi2022excitonic}. In this work, we elucidate the quantum geometric properties of interacting systems in the context of excitons, and demonstrate that these properties are manifested in the experimentally observable non-linear optical response. We show that the quantum geometry of the correlated electron-hole state cannot be simply attributed to the quantum geometry of the constituent single-particle states. There is an additional quantum many-body geometric term that involves the phase coherence of the many-body state. For the shift current, a second-order optical response, the quantum many-body geometric term may be interpreted as a many-body analogue of the Berry phase. We emphasise that the quantum many-body geometry we elucidate here is distinct from previous discussions on the role of quantum geometry of single-particle states in non-linear optics~\cite{ahn2020low}, which have been the basis for the interpretation of recent experiments~\cite{sodemann2015quantum,xu2018electrically,ma2019observation,beaulieu2024berry}. Using first principles calculations within the $GW$-Bethe-Salpeter Equation (BSE) approach, we demonstrate that this quantum many-body geometric term is important for quantitative comparison with experiment for the low-temperature shift current conductivity in bulk CdS, and furthermore, significantly enhances the non-linear optical response in systems with strong excitonic interactions. \\
We begin our discussion with the shift current, which is a direct current arising from the second-order light-matter interaction. Within the independent/single particle approximation (IPA), the shift current conductivity in the direction $\alpha$ is given by:~\cite{young2012first,sipe2000second}
\begin{equation}\label{sc_ipa}
\sigma^{\alpha,\text{IPA shift}}_{\beta\beta} = \mathcal{\pi C}\sum_{nm\vb{k}}\mathcal{R}^{\alpha}_{nm\vb{k}}\abs{r^{\beta}_{nm\vb{k}}}^2 f_{mn\vb{k}} \delta(\omega_{mn\vb{k}}-\omega),
\end{equation}
where
\begin{equation}\label{sv_ipa}
\mathcal{R}^{\alpha}_{nm\vb{k}}=\xi^{\alpha}_{nn\vb{k}}-\xi^{\alpha}_{mm\vb{k}} + \pdv{\phi_{nm\vb{k}}}{k^{\alpha}}
\end{equation}
is known as the shift vector. Here, $\xi^{\alpha}_{nn\vb{k}}\equiv i\braket{u_{n\bf{k}}}{\grad_{\bf{k}}u_{n\bf{k}}}$ is the Berry connection for Bloch states and $\phi_{nm\vb{k}}$ is the phase of the dipole matrix element with $r^\beta_{nm\vb{k}} = \abs{r^\beta_{nm\vb{k}}}e^{-i\phi_{nm\vb{k}}}$ for light polarised along direction $\beta$. The IPA shift vector describes a real-space shift of the charge centre of the photo-excited electron, and $\xi^{\alpha}_{nn\vb{k}}$ is involved because the position operator is defined by~\cite{aversa1995nonlinear} 
\begin{equation}\label{pos_op}
	\mel{n\vb{k}}{\hat{\vb{r}}}{m\vb{k}'} = \vb{\xi}_{nm\vb{k}}\delta(\vb{k}-\vb{k}')+\delta_{nm}i\grad_{\vb{k}}\delta(\vb{k}-\vb{k}').
\end{equation}
The term $\pdv{\phi_{nm\vb{k}}}{k^{\alpha}}$ ensures that the shift vector is invariant to changes in the gauge of the Bloch states ($\ket{n\vb{k}}\rightarrow e^{-i\theta_{n\vb{k}}}\ket{n\vb{k}}$)~\cite{}.
\\
To understand the interplay between quantum geometry and interactions in the presence of excitonic interactions, we evaluate the expression for the shift current conductivity using a similar approach to Ref.~\parencite{sipe2000second} for the IPA, but in the basis of many-body excited states. These states are obtained by solving the BSE within the Tamm-Dancoff approximation~\cite{rohlfing2000electron}:
\begin{equation}\label{bse} 
(\varepsilon_{c\vb{k}}- \varepsilon_{v\vb{k}}) A^S_{vc\vb{k}} + \sum_{v'c'\vb{k}'}K^{eh}_{vc\vb{k}, v'c'\vb{k}'} A^S_{v'c'\vb{k}'} = (E_S-E_0) A^S_{vc\vb{k}},
\end{equation} 
Here, $\varepsilon$ are the quasiparticle energies obtained using the $GW$ approximation, $K^{eh}$ is the electron-hole interaction kernel, and $(E_S-E_0)$ are the excitation energies. The corresponding correlated exciton states are given as $\ket{S} = \sum_{vc\vb{k}}A^S_{vc\vb{k}}\hat{a}^\dagger_{c\vb{k}}\hat{a}_{v\vb{k}}\ket{0}$   
where $\hat{a}^\dagger$ and $\hat{a}$ are the electron creation and annihilation operators, $\ket{0}$ is the ground-state Slater determinant and $A^S_{vc\vb{k}}$ is the exciton envelope function. The resulting shift current conductivity is given by (see Supporting Information)
\begin{equation}\label{sc_r}
\sigma^{\alpha,\text{shift}}_{\beta\beta} = \mathcal{\pi C}\sum_{S}R^{\alpha}_{S0}\abs{r^{\beta}_{0S}}^2 \delta(\Omega_{S}-\Omega_0-\omega),  
    \end{equation}
where $R^{\alpha}_{S0}$ is given by
	\begin{equation}\label{sv_bse}
	R^{\alpha}_{S0} = \mel{S}{\hat{r}^\alpha}{S} - \mel{0}{\hat{r}^\alpha}{0}
	\end{equation}
with $\omega>0$, $\mathcal{C} = \frac{e^3}{\hbar^2V_{crys}}<0$ and $\hat{r}^\alpha = \sum_{nm\vb{k}\vb{k}'} r^\alpha_{nm}(\vb{k}, \vb{k}')\hat{a}^{\dagger}_{n\vb{k}}\hat{a}_{m\vb{k}'}$ . It is natural to identify $R^{\alpha}_{S0}$ as the many-body shift vector. A similar expression for the many-body shift current conductivity has been presented in Ref.~\parencite{resta2024geometrical} where the many-body shift vector is defined in terms of derivatives with respect to a generalised parameter for the many-body Hamiltonian. We note that the many-body shift vector depends only on the initial and final expectation values of $\hat{r}$ and is independent of the direction of polarisation of light ($\beta$). \\
In this work, $R^{\alpha}_{S0}$ is evaluated in terms of the many-body exciton state as:
\begin{align}\label{R_S01}
\begin{split} 
       R^{\alpha}_{S0} &= \mel{S}{\hat{r}^\alpha}{S} - \mel{0}{\hat{r}^\alpha}{0}\\&= \sum_{vcv'c'\vb{k}}\left(A^{S^*}_{vc'\vb{k}}A^{S}_{vc\vb{k}}\xi^{\alpha}_{c'c\vb{k}} - A^{S^*}_{v'c\vb{k}}A^{S}_{vc\vb{k}}\xi^{\alpha}_{vv'\vb{k}}\right)
+\sum_{\substack{vc\vb{k}}}\left(iA^{S^*}_{vc\vb{k}}\pdv{A^{S}_{vc\vb{k}}}{k^{\alpha}_v} - iA^{S}_{vc\vb{k}}\pdv{A^{S^*}_{vc\vb{k}}}{k^{\alpha}_c} \right).
\end{split}
\end{align}
It is instructive to compare $R^{\alpha}_{S0}$ with the IPA shift vector to build physical intuition. In the IPA limit, where $A^{S}_{vc\vb{k}}$ is non-zero for a particular $\{vc\vb{k}\}$ index and 0 otherwise, the first sum in Eq.~\ref{R_S01} reduces to the first two terms in the IPA shift vector, $(\xi^{\alpha}_{nn\vb{k}}-\xi^{\alpha}_{mm\vb{k}})$, modulo phase factors. The second sum, $\sum_{\substack{vc\vb{k}}}\left(iA^{S^*}_{vc\vb{k}}\pdv{A^{S}_{vc\vb{k}}}{k^{\alpha}_v} - iA^{S}_{vc\vb{k}}\pdv{A^{S^*}_{vc\vb{k}}}{k^{\alpha}_c} \right)$, plays the same role as $\pdv{\phi_{nm\vb{k}}}{k^{\alpha}}$ in the IPA shift vector; any changes in phases of the Bloch states are reflected in changes in phases for $A^{S}_{vc\vb{k}}$ through the BSE, and this term ensures gauge-invariance of the many-body shift vector (see Supporting Information). It is this second sum that contains the quantum many-body geometric terms.  \\
Considering only momentum-direct optical transitions in this analysis, so that $k_v = k_c = k$, the second sum in Eq.~\ref{R_S01} can be written as:
\begin{equation}\label{geo_term2}
\begin{split}&\sum_{vc\vb{k}}\left(iA^{S^*}_{vc\vb{k}}\pdv{A^{S}_{vc\vb{k}}}{k^{\alpha}} - iA^{S}_{vc\vb{k}}\pdv{A^{S^*}_{vc\vb{k}}}{k^{\alpha}}\right) = \sum_{vc\vb{k}}2i\text{Im}\left(A^{S^*}_{vc\vb{k}}\pdv{A^{S}_{vck}}{k^{\alpha}}\right)\\&=-\frac{1}{\delta k^\alpha}\sum_{vc\vb{k}}2i\abs{A^S_{vc\vb{k}}}^2\text{Im}\left[\text{ln}\left(\tilde{A}^{S*}_{vc\vb{k}}\tilde{A}^S_{vc\vb{k}+\delta k^\alpha}\right)\right]
	\end{split}
	\end{equation}
with $\tilde{A}^{S*}_{vc\vb{k}} = {A}^{S*}_{vc\vb{k}}/\abs{A^S_{vc\vb{k}}}$ and $\tilde{A}^S_{vc\vb{k}+\delta k^\alpha} = {A}^S_{vc\vb{k}+\delta k^\alpha}/\abs{A^S_{vc\vb{k}}}$. Here, $\text{Im}\left[\text{ln}\left(\tilde{A}^{S*}_{vc\vb{k}}\tilde{A}^S_{vc\vb{k}+\delta k^\alpha}\right)\right]$ describes the phase difference of the envelope function at two points in the Brillouin zone (BZ). This is analogous to the Berry connection of the Bloch states which measures the phase difference between two Bloch states indexed by wavevectors that are infinitesimally spaced apart in the BZ. This can be seen from the definition of the Berry connection~\cite{Vanderbilt_2018}
    \begin{equation}
    \xi^{\alpha}_{nn\vb{k}}\equiv i\braket{u_{n\bf{k}}}{\grad_{\bf{k}}u_{n\bf{k}}}=\frac{1}{\delta k^\alpha}i\ln \braket{u_{n\bf{k}}}{u_{n\bf{k}+\delta k^\alpha}} = -\frac{1}{\delta k^\alpha}\text{Im}\left[\ln \braket{u_{n\bf{k}}}{u_{n\bf{k}+\delta k^\alpha}}\right].
    \end{equation}
Just like how the Berry phase is a sum over a closed loop of the Berry connection for single-particle Bloch states, the quantum many-body geometric term here can be seen as a series of closed-loop summations of the Berry connection analogue for the correlated exciton envelope function. We illustrate this in Figure~\ref{geometric}.  Using a two dimensional BZ as an example, this sum can be broken down into a series of summations along straight lines across the BZ as depicted in Figure~\ref{geometric}(a), and the direction of these lines is given by the component $\alpha$ for the shift vector $R^\alpha_{S0}$. Due to the periodicity of the lattice, these lines can be thought of as loops around a closed surface, by folding the BZ and forming a cylinder (Figure~\ref{geometric}(b)). The quantum many-body geometric term is then a series of closed-loop summations involving the overlaps of the exciton envelope function, and accumulates a phase upon traversing the loops.\\
This quantum many-body geometric part of the shift vector can also be interpreted simply as resulting from variations over the BZ of the complex phase of the exciton envelope function. Writing $A^S_{vck} = \abs{A^S_{vck}}e^{-i\Gamma^S_{vc}(\vb{k})}$, we have
\begin{equation}\label{geo_term1}
	\begin{split}
&\sum_{vc\vb{k}}\left(iA^{S^*}_{vc\vb{k}}\pdv{A^{S}_{vc\vb{k}}}{k^{\alpha}} - iA^{S}_{vc\vb{k}}\pdv{A^{S^*}_{vc\vb{k}}}{k^{\alpha}}\right) =  2\sum_{vc\vb{k}} \pdv{\Gamma^S_{vc}(\vb{k})}{k^\alpha}\abs{A^S_{vc\vb{k}}}^2, 
	\end{split}
	\end{equation}
i.e. the term is a weighted sum of the derivative of the phase of $A^S_{vck}$, summed over the BZ and over all valence and conduction band pairs. For a real-valued continuous envelope function, $\Gamma_{vc}(\vb{k})$ is a constant value of either $\pi$ or 0 and the term vanishes. The quantum many-body geometric term therefore captures the complex phase coherence of the correlated exciton state and does not have a direct analogy in the IPA.\\
We will now proceed to quantify using first-principles $GW$-BSE calculations the implications of the quantum many-body geometric terms. It is helpful to rewrite $R^\alpha_{S0}$ in terms of the IPA shift vector, which leads to a partition of $R^\alpha_{S0}$ into three gauge-invariant terms.
\begin{equation}\label{sc_terms}
    \begin{split}
        R^\alpha_{S0}&= \underbrace{\sum_{\substack{v\vb{k}\\c\neq c'}}A^{S^*}_{vc'\vb{k}}A^{S}_{vc\vb{k}}\xi^{\alpha}_{c'c\vb{k}} - \sum_{\substack{c\vb{k}\\v\neq v'}}A^{S^*}_{v'c\vb{k}}A^{S}_{vc\vb{k}}\xi^{\alpha}_{vv'\vb{k}}}_{\text{Term A}}+\underbrace{\sum_{\substack{vc\vb{k}}}\abs{A^{S}_{vc\vb{k}}}^2\mathcal{R}^\alpha_{cv\vb{k}}}_{\text{Term B}}\\&
+\underbrace{\sum_{\substack{vc\vb{k}}}\left(iA^{S^*}_{vc\vb{k}}\pdv{A^{S}_{vc\vb{k}}}{k^{\alpha}_v} - iA^{S}_{vc\vb{k}}\pdv{A^{S^*}_{vc\vb{k}}}{k^{\alpha}_c} -\abs{A^{S}_{vc\vb{k}}}^2\pdv{\phi_{cv\vb{k}}}{k^{\alpha}}\right).}_{\text{Term C}}
    \end{split}
\end{equation}
Term A can be interpreted as interband transitions within the conduction and valence band manifolds; this term is absent in the IPA and we have found numerically that this term tends to be one to two orders of magnitude smaller than the other terms. Term B can be described as a renormalized IPA shift vector. Term C contains the quantum many-body geometric effects and we shall henceforth refer to Term C as the quantum many-body geometric term. Term C is gauge-invariant and the contribution of Term C to the shift current conductivity will be explicitly computed below. \\
\begin{figure}
    \includegraphics[width = \textwidth]{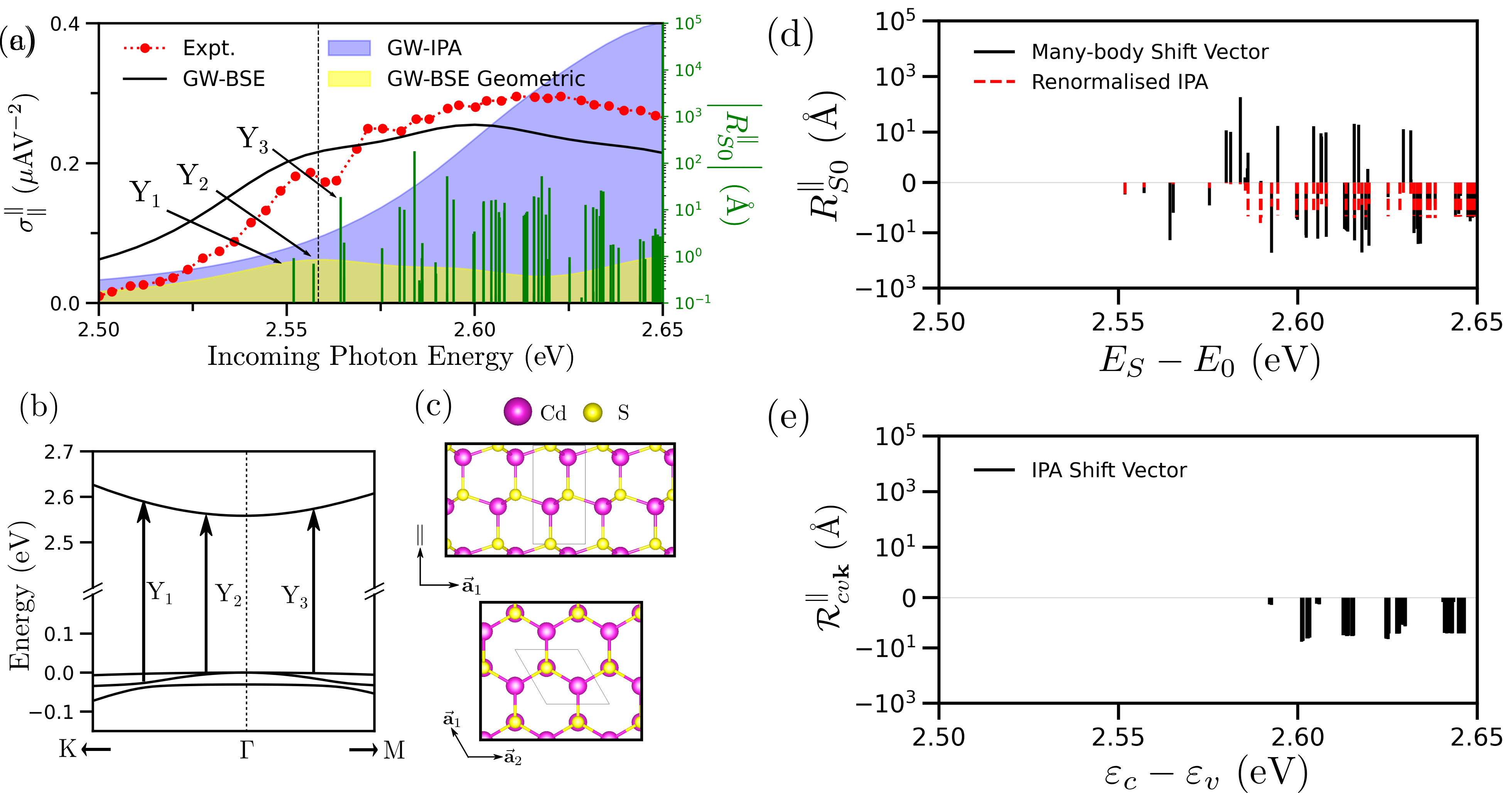}
    \caption{(a) Shift current conductivity (left axis) and magnitude of the many-body shift vectors (right axis) for bulk CdS. The shift current conductivity is computed with $GW$-BSE (black curve) and $GW$-IPA (blue shaded region). Experimental data is taken from Ref.~\parencite{sotome2021terahertz}. $\parallel$ refers to the direction parallel to the polar axis. The vertical dashed line represents the quasiparticle band gap. The yellow shaded region indicates the contribution of Term C in Eq.~\ref{sc_terms} to the shift current conductivity. The magnitude of the many-body shift vector is also plotted (in green) as a function of exciton energy, to understand the contribution of individual excitons to the shift current. (b) $GW$ quasiparticle bandstructure around $\Gamma$ for CdS, with vertical arrows indicating the interband transitions for the three excitons labelled in (a). The optical absorption spectra compared with experiment is shown in Fig. S1. (c) Atomic structure of CdS. We consider the shift current component along the $\parallel$ axis. (d) Many-body shift vector along the $\parallel$ direction shown as a function of the exciton energy. The renormalized IPA contribution (Term B in Eq.~\ref{sc_terms}) is shown as dashed red lines. (e) The IPA shift vector along the $\parallel$ direction as a function of the interband transition energies.}
    \label{sc_cds}
\end{figure}
We first validate our approach by computing the shift current 
conductivity for bulk wurtzite CdS and compare our results to experimental data measured at 2 Kelvins~\cite{sotome2021terahertz}. The low temperature for the experimental measurement ensures that the static second-order optical response is dominated by the shift current and not by ballistic currents~\cite{sturman2020ballistic}. Our predicted shift current conductivity compares reasonably well with experiment (black curve and red dotted line in Figure~\ref{sc_cds}(a)). The measured shift current exhibits a peak beneath the quasiparticle band-gap of CdS; this peak can only be accounted for when electron-hole interactions are considered. We analyse the contribution of these peaks by evaluating the magnitude of the many-body shift vectors for all optically-active excitons. The below-band-gap peak can be attributed primarily to three excitons whose dominant transitions are shown in Figure~\ref{sc_cds}(b). The IPA shift current conductivity, computed without electron-hole interactions, is shaded in blue in Figure~\ref{sc_cds}(a). Unlike for 2D materials~\cite{chan2021giant,li2021enhanced,nakamura2024strongly,ruan2024exciton,abdelwahab2022giant,abdelwahab2023highly,wu2022data,jia2019niobium}, the shift current is not significantly enhanced by excitonic interactions for bulk CdS. However, as discussed above, signatures of excitonic effects are evident in the experimental measurement. We have also isolated the contribution of Term C (the quantum many-body geometric term) of the shift vector to the total many-body shift current conductivity; this is shown in yellow in Figure~\ref{sc_cds}(a). The quantum many-body geometric contribution is important for quantiative agreement with experiment. One can also observe from Figure~\ref{sc_cds}(d) that the many-body shift vector is quite different from the renormlised IPA shift vector (Term B), further underscoring the importance of the quantum many-body geometric term.\\

\begin{figure}
    \centering
\includegraphics[width=1\textwidth]{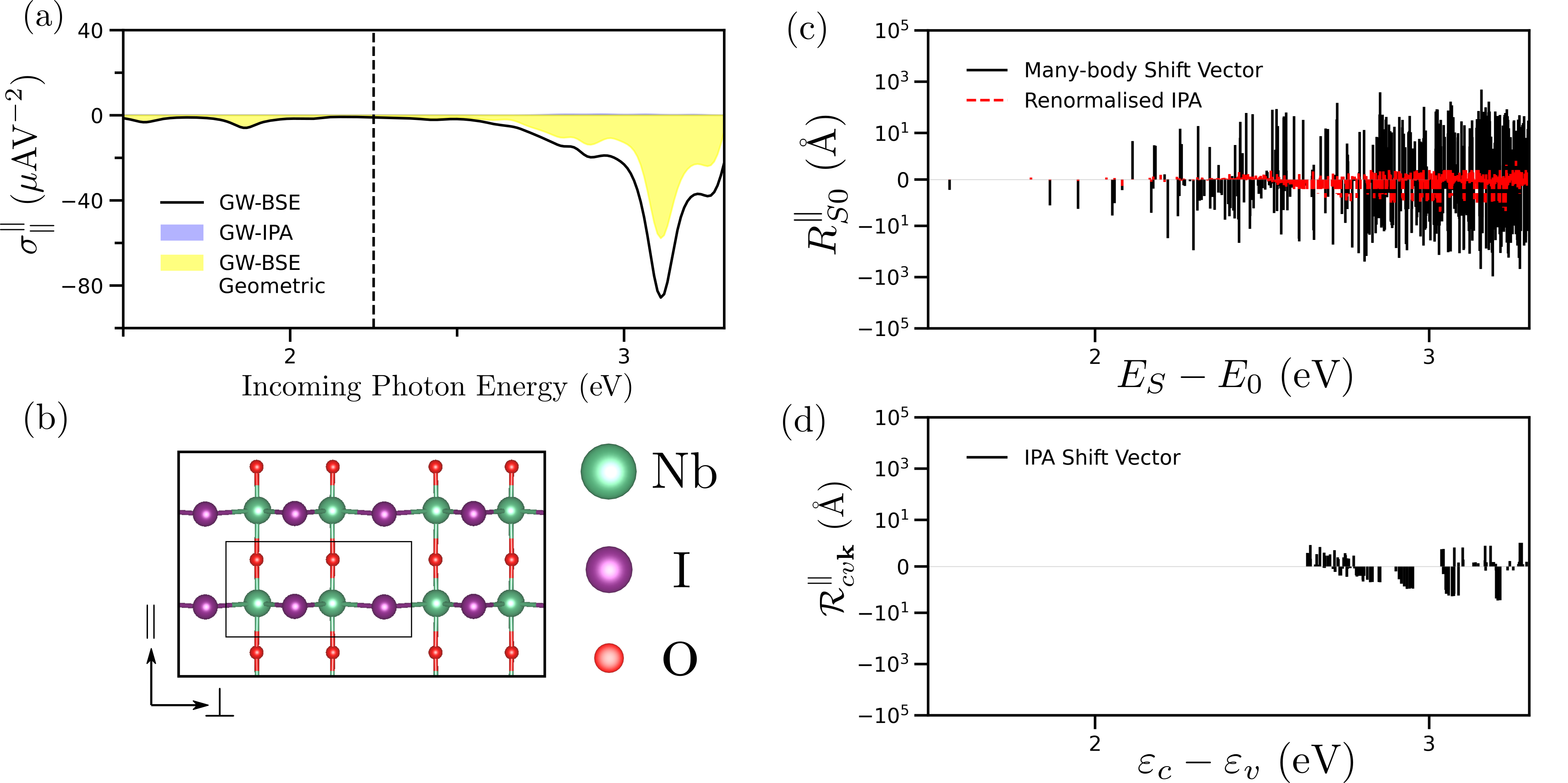}
    \caption{(a) Shift current conductivity for monolayer NbOI$_2$.  The shift current conductivity is computed with $GW$-BSE (black curve) and $GW$-IPA (blue shaded region). The vertical dashed line represents the quasiparticle band gap. The IPA shift current conductivity (between 2.5 eV to 3.3 eV) is roughly two orders of magnitude smaller compared to the many-body shift current conductivity, with the largest value of around 0.51 $\mu$AV$^{-2}$ at around 2.9 eV. The many-body shift current conductivity on the other hand has a maximum of around -80 $\mu$AV$^{-2}$ at around 3.1 eV. The optical absorption spectra compared with experiment is shown in Fig. S2. (b) Atomic structure for monolayer NbOI$_2$. The polar ($\parallel$) and non-polar directions ($\perp$) are indicated. We consider the shift current component along the $\parallel$ axis. (c) Many-body shift vector along the $\parallel$ direction shown as a function of the exciton energy. The renormalized IPA contribution (Term B in Eq.~\ref{sc_terms}) is shown as dashed red lines. (d) The IPA shift vector along the $\parallel$ direction as a function of the interband transition energies.}
    \label{nboi2_sc}
\end{figure}
Next, we present our results for 2D NbOI$_2$, a ferroelectric material that has attracted significant attention for applications in non-linear optics~\cite{abdelwahab2022giant,xuan2024exciton,AdvMaterNbOI2} and piezoelectrics~\cite{wu2022data}. It has been demonstrated that the properties of NbOI$_2$ do not depend significantly on thickness~\cite{jia2019niobium,abdelwahab2023highly} and electron-hole interactions are important not just in the monolayer, but also for the layered bulk system~\cite{abdelwahab2022giant}. Electron-hole interactions enhance the shift current conductivity for monolayer NbOI$_2$ by orders of magnitude (see Figure~\ref{nboi2_sc}(a) and associated caption). In this case, the quantum many-body geometric term (Term C in Eq.~\ref{sc_terms}) is shown to account for the majority of the interaction-enhanced shift current conductivity (yellow shading in Figure~\ref{nboi2_sc}(a)), demonstrating a clear advantage induced by quantum many-body geometry in enhancing the shift current response. This conclusion is corroborated by the shift vector plots in Figure~\ref{nboi2_sc}(c-d) where it can be observed that the many-body shift vector is in general about two orders of magnitude longer than the IPA shift vector or the renormalised IPA term (Term B in Eq.~\ref{sc_terms}).\\

\begin{figure}
\centering\includegraphics[width=0.7\textwidth]{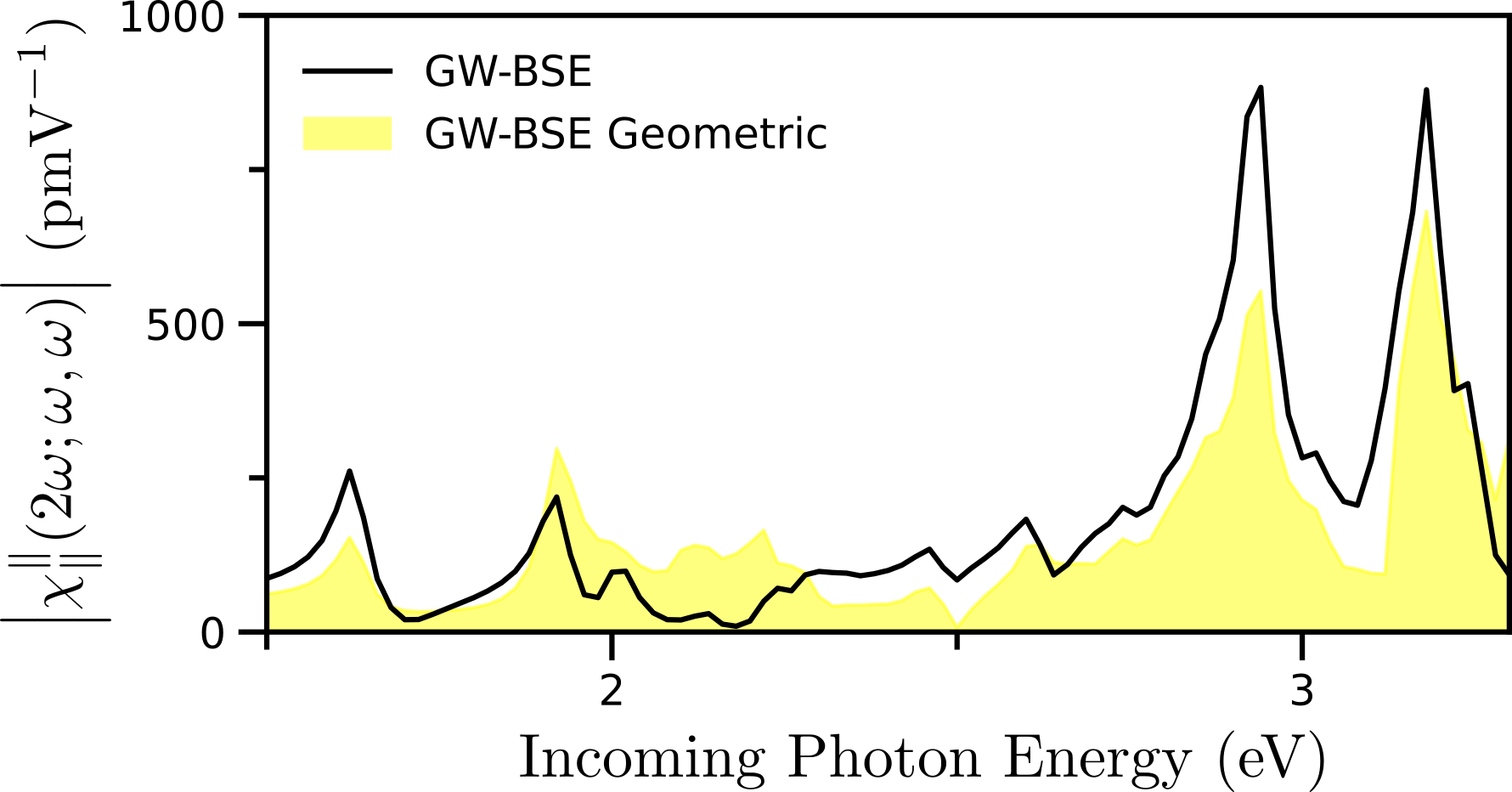}
    \caption{Magnitude of the susceptibility tensor corresponding to second harmonic generation for monolayer NbOI$_2$; both the incident and emitted light are polarised along the $\parallel$ axis (see Figure~\ref{nboi2_sc}). The yellow shaded area represents the contribution of the geometric term to the total susceptibility (black solid lines).}
    \label{nboi2_shg}
\end{figure}

Interaction-enhanced shift currents have been described in recent literature, for low-dimensional systems where excitionic effects are significant~\cite{li2021enhanced,chan2021giant,nakamura2024strongly,ruan2024exciton}. Do we expect the quantum many-body geometric term to enhance the shift current in other systems with strong excitonic interactions? We first note that it has been suggested~\cite{chan2021giant} that the excitonic enhancement in the shift current arises primarily to increased oscillator strengths when electron-hole interactions are taken into account. While increased oscillator strengths do give rise to a larger shift current, our analysis for NbOI$_2$ shows that the increase in the magnitude of the shift vector when electron-hole interactions are considered is much larger than the increase in oscillator strength (see also Figure S2). Furthermore, the oscillator strengths are governed by a sum rule~\cite{BASSANI2005200} so that it is unlikely for oscillator strengths to be enhanced over all exciton states; thus, enhancements in oscillator strength are not likely to be responsible for the overall enhancement observed for non-linear optical responses. We therefore expect that the excitonic-enhancement of the shift current arises primarily from the excitonic-enhancement in the shift vector. Returning to Eq.~\ref{sc_terms}, it is easy to see that any excitonic enhancement in the shift vector comes from the quantum many-body geometric term - Term C. Specifically, Term A is negligible, while Term B is of the same order as the IPA shift vectors, since the exciton envelope function is normalised to one (ie. $\sum_{vc\vb{k}}\abs{A^S_{vc\vb{k}}}^2 = 1$). Another way to see intuitively that the quantum many-body geometric term enhances the shift vector is to consider the real-space representation. We can write
\begin{equation}\label{sv_e-h}
\begin{split}
    R^\alpha_{S0} &= \mel{S}{r^\alpha}{S} - \mel{0}{r^\alpha}{0}\\ &= \int d\vb{r}_e d\vb{r}_h \Psi^{S*}(\vb{r}_e, \vb{r}_h)[r^\alpha_e-r^\alpha_h]\Psi^S(\vb{r}_e, \vb{r}_h),
\end{split}
\end{equation}
where the exciton wavefunction is
\begin{equation}\label{Psi_S}
 \Psi^S(\vb{r}_e, \vb{r}_h) = \sum_{vc\vb{k}}A^S_{vc\vb{k}}\psi^*_{v\vb{k}}(\vb{r}_h)\psi_{c\vb{k}}(\vb{r}_e)  
\end{equation}
 and the integral is taken over all space. Writing $\vb{r}_e$ as $(\vb{r}_e\rvert_{\vb{r}_e \in \Omega}+\vb{R}_e)$ (and likewise for $\vb{r}_h$), where $\Omega$ refers to a chosen unit cell, and $\vb{R}_e$ a lattice vector, we arrive at: 
\begin{equation}\label{R_S0_real}
    \begin{split}
        R^\alpha_{S0} &= \sum_{vc\vb{k}}A^{S*}_{vc\vb{k}} A^S_{vc\vb{k}}\frac{1}{\Omega}\int_\Omega d\vb{r}_e u^*_{c\vb{k}}r^\alpha_eu_{c\vb{k}} - \sum_{cv\vb{k}}A^{S*}_{vc\vb{k}} A^S_{vc\vb{k}}\frac{1}{\Omega}\int_\Omega d\vb{r}_h u^*_{v\vb{k}}r^\alpha_h u_{v\vb{k}}\\
        &+\sum_{\vb{R}} \vb{R}_e^\alpha \int_{\Omega(\vb{R}_e)}d\vb{r}_e\rho^S(\vb{r}_e) - \sum_{\vb{R}}\vb{R}^\alpha_h\int_{\Omega(\vb{R}_h)}d\vb{r}_h\rho^S(\vb{r}_h).
    \end{split}
\end{equation}
Here the integrals are performed over the chosen unit cell $\Omega$ in the first line, while $\Omega(\vb{R})$ refers to the unit cell obtained by translating the chosen unit cell by the lattice vector $\vb{R}$, and the sums in the second line are taken over all lattice vectors in the crystal. We have defined an electron(hole) density to be $\rho^S(\vb{r}_{e(h)}) = \int d\vb{r}_{h(e)}\abs{\Psi^S(\vb{r}_e, \vb{r}_h)}^2$ which describes the spatial distribution of the electron(hole) averaged over all possible hole(electron) positions.
The first line of Eq.~\ref{R_S0_real} resembles the renormalized IPA shift vector. The second line can thus be interpreted as the geometric term, which can be interpreted as a shift in the electron density due to the excitation, averaged over all lattice sites. The quantum many-body geometric term thus gives a ``global'' or coarser view of the electron-hole separation within the crystal when the exciton is formed, and intuitively, it can be understood that this component of the shift vector spans over multiple unit cells and can dominate the shift vector. \\
In the above, we have demonstrated that the interplay between quantum geometry and interactions leads to an additional quantum many-body geometric term that gives rise to effects that cannot be explained by a superposition of quantum geometric effects of single-particle states. In particular, the quantum many-body geometric term in the shift vector can be interpreted as a geometric phase arising from the complex phase dependence of the exciton envelope function, and the quantum geometric term enhances the shift current. We note that the quantum many-body geometric term inherently arises from the matrix element $r^\alpha_{S'S}$, which shows up in the general non-linear optical susceptibility~\cite{ruan2024exciton}. For $S\neq S'$, the quantum many-body geometric term involves $\left(iA^{S'^*}_{vc\vb{k}}\pdv{A^{S}_{vc\vb{k}}}{k^{\alpha}_v} - iA^{S}_{vc\vb{k}}\pdv{A^{S'^*}_{vc\vb{k}}}{k^{\alpha}_c}\right)$, which resembles an off-diagonal multi-band Berry connection term for Bloch states~\cite{Vanderbilt_2018,manini2000off}. In previous work~\cite{xuan2024exciton}, we showed that excitonic interactions enhance by orders of magnitude the non-linear optical response corresponding second harmonic generation (SHG) in NbOI$_2$, and are essential for explaining the experimental polar plots. In Figure~\ref{nboi2_shg}, we show in black the magnitude of the susceptibility tensor $\chi^{(2)}(2\omega;\omega,\omega)$ corresponding to SHG for monolayer NbOI$_2$, as computed in Ref.~\parencite{xuan2024exciton}. Isolating the quantum many-body geometric contribution to $\lvert\chi^{(2)}(2\omega;\omega,\omega)\rvert$ (see Supporting Information), we see that the quantum many-body geometric term accounts for most of $\lvert\chi^{(2)}(2\omega;\omega,\omega)\rvert$ (yellow shading in Figure~\ref{nboi2_shg}). This clearly demonstrates that the quantum many-body geometric term can also lead to an enhancement in the general non-linear optical response for systems with strong excitonic interactions.  
\\
In conclusion, we have demonstrated that the quantum geometry of interacting particles cannot be simply described in terms of the quantum geometry of the constituent single particles. We show that there are additional quantum geometric effects that involve the complex phase of the correlated state and in the case of the shift current, can be interpreted as the accumulation of a geometric phase arising from the exciton envelope function. This quantum many-body geometric term significantly enhances the non-linear optical response in systems with strong excitonic interactions - an effect that we describe as a quantum geometric advantage of the correlated exciton state. Our work demonstrates that excitonic devices are a desirable platform to study the geometric properties of correlated states of interacting particles through nonlinear optics.

\section{Acknowledgements}
This work is supported by NUS and the National Research Foundation (NRF), Singapore, under the NRF medium-sized centre programme, as well as by A*STAR, Singapore under Project ID M24N7c0095.
Calculations were performed on the National Supercomputing Centre, Singapore. ML also thanks LQN Cheng and YL Kwok for support and discussions.

\section{Computational Methods}
The first principles $GW$-BSE calculations are performed with the BerkeleyGW code~\cite{deslippe2012berkeleygw}, with the mean-field starting point obtained with density functional theory calculations performed using QuantumESPRESSO~\cite{giannozzi2009quantum,giannozzi2017advanced}. Optimized norm-conversing Vanderbilt (ONCV)~\cite{hamann2013optimized} pseudopotentials were used~\cite{van2018pseudodojo}. The non-linear optical properties in the many-body regime are computed using the exciton envelope function and excitation energies obtained by solving the BSE in the BerkeleyGW code. \\
To compute the shift current conductivity, we used the velocity-gauge expression
\begin{equation}\label{sc} 
    \begin{split}
\sigma^{\alpha,\text{shift}}_{\beta\beta} &= 	\frac{-i\mathcal{C}}{m^2\omega^2}\sum_{SS'}\frac{v^{\alpha}_{0S'}p^{\beta}_{S'S}p^{\beta}_{S0}}{\Omega_0-\Omega_{S'}}\delta(\Omega_{S}-\Omega_0-\omega)
    +\frac{i\mathcal{C}}{m^2\omega^2}\sum_{\Omega_S\neq\Omega_{S'}}\frac{v^{\alpha}_{S'S}p^{\beta}_{S0}p^{\beta}_{0S'}}{\Omega_{S'}-\Omega_{S}}\delta(\Omega_{S}-\Omega_0-\omega).
    \end{split}
\end{equation} 
The many-body shift vector was computed using a sum rule
\begin{equation}\label{commutator}
    R^{\alpha}_{S0}=\frac{1}{p^\beta_{0S}}\left(\sum_{S'\neq 0} r_{0S'}^{\alpha}p_{S'S}^{\beta} - \sum_{S'\neq S}p^\beta_{0S'}r^{\alpha}_{S'S}\right).
\end{equation}
To compute the quantum many-body geometric contribution to the shift current conductivity, we evaluate the shift current conductivity using Eq.~\ref{sc_r} with only Terms A and B in $R^{\alpha}_{S0}$ (see Eq.~\ref{sc_terms}) and then subtract this from the total shift current conductivity. The IPA shift vector in Term B is evaluated using an IPA sum rule~\cite{aversa1995nonlinear,rangel2017large}.\\
For the non-linear optical susceptibility corresponding to SHG, we compute the geometric contribution by evaluating Term C in $r^{\alpha}_{S'S}$ below:
\begin{equation} \label{rssp} 
    \begin{split} 
       r^{\alpha}_{S'S} &= \underbrace{\sum_{\substack{v\vb{k}\\c\neq c'}}A^{S'^*}_{vc'\vb{k}}A^{S}_{vc\vb{k}}\xi^{\alpha}_{c'c\vb{k}} - \sum_{\substack{c\vb{k}\\v\neq v'}}A^{S'^*}_{v'c\vb{k}}A^{S}_{vc\vb{k}}\xi^{\alpha}_{vv'\vb{k}}}_{\text{Term A}}+
\underbrace{\sum_{\substack{vc\vb{k}}}A^{S'^*}_{vc\vb{k}}A^{S}_{vc\vb{k}}\mathcal{R}^\alpha_{cv\vb{k}}}_{\text{Term B}}\\&+\underbrace{iA^{S'^*}_{vc\vb{k}}\pdv{A^{S}_{vc\vb{k}}}{k^{\alpha}_v} - iA^{S}_{vc\vb{k}}\pdv{A^{S'^*}_{vc\vb{k}}}{k^{\alpha}_c}-A^{S'^*}_{vc\vb{k}}A^{S}_{vc\vb{k}}\pdv{\phi_{cv\vb{k}}}{k^\alpha}}_{\text{Term C}}
      + r^{\alpha}_{00}\delta_{S'S},
    \end{split} 
    \end{equation}
and use Term C in place of $r^{\alpha}_{S'S}$ in Eq. 1 of Ref.~\parencite{ruan2024exciton}.\\
Details of the numerical parameters used in the calculations are provided in the Supporting Information. 
\section{Data Availability}
The data presented in this study are available from the corresponding author upon request.
\bibliography{ref.bib}
\section{Author Contributions}
This work was performed by ML under the supervision of SYQ. FX contributed to early discussions with ML and SYQ.
\section{Competing Interests}
The authors declare no competing interests.
\end{document}